\documentclass[aps,prd,twocolumn,floatfix,nofootinbib,showpacs,superscriptaddress]{revtex4-1}
\pdfoutput=1
\usepackage{amsmath,amsfonts,amssymb,bm,ulem}

\usepackage{graphicx}
\usepackage{color}
\usepackage{url} 
\usepackage{threeparttable}
\usepackage{makecell}

\definecolor{purple}{rgb}{0.1,0.5,0.4}

\usepackage[
pdfauthor={derajan},
pdfstartview=XYZ,
bookmarks=true,
colorlinks=true,
linkcolor=blue,
urlcolor=blue,
citecolor=blue,
pdftex,
bookmarks=true,
linktocpage=true,   % makes the page number as hyperlink in table of content
hyperindex=true
]{hyperref} 

\begin{document}

\title{Cosmological angular momentum from quantum rotation}

\author{Bo-Qiang Lu} 
\email{bqlu@zjhu.edu.cn}
\affiliation{School of Science, Huzhou University, Huzhou, Zhejiang 313000, P. R. China}

\begin{abstract}
The origin of cosmic angular momentum is a fundamental question in structure formation. We propose a novel mechanism that generates spatial angular momentum directly from quantum fluctuations during inflation. A spectator complex scalar field with global U(1) symmetry stores internal angular momentum via field-space rotation. Inflationary perturbations create spatial gradients that, upon horizon re-entry, couple to the background charge density and source a bulk momentum flow. During nonspherical gravitational collapse, this flow converts into net angular momentum. For primordial black holes forming from such collapse, the dimensionless spin can reach \(\chi\sim 0.1-1\) when the small-scale power spectrum is enhanced to produce detectable abundances—far exceeding tidal torque theory predictions. This establishes a testable link between inflation, primordial perturbations, and black hole spin distributions accessible to gravitational-wave observations.

\end{abstract}
\pacs{}
\maketitle

% \section{Introduction}\label{sec:intro}
\textbf{Introduction.}
The universe is replete with rotation, from swirling galaxies to spinning black holes (BHs). Understanding the origin of this angular momentum is crucial for a complete theory of structure formation. The standard framework for explaining the spin of galaxies and dark matter (DM) halos is the classical tidal torque theory (TTT)~\cite{Peebles:1969jm,White:1984uf,Catelan:1996hv,Jones:2008fv}. TTT posits that a protohalo, embedded in a tidal gravitational field generated by neighboring large-scale structures, acquires angular momentum due to a misalignment between its moment of inertia tensor and the external tidal tensor. This theory successfully predicts typical spin parameters for DM halos, $\chi \sim 0.01-0.1$, largely independent of mass, a result confirmed by modern cosmological simulations~\cite{Boylan-Kolchin:2009alo,Pillepich:2017jle}.

However, TTT has inherent limitations. It describes the conversion of existing asymmetries into angular momentum, not its fundamental creation. The ultimate origin of the necessary initial density and velocity perturbations is traced back to quantum fluctuations during inflation. Furthermore, TTT struggles to explain the extremely high spins inferred for some supermassive BHs without invoking subsequent processes like major mergers or extensive gas accretion.
%==========================================================================

Primordial black holes (PBHs), which could form from the collapse of large-amplitude density perturbations in the early universe~\cite{Hawking:1971ei,Carr:1974nx}, are compelling DM candidates and potential sources for gravitational wave (GW) events observed by LIGO-Virgo-KAGRA (LVK)~\cite{Bird:2016dcv,Sasaki:2016jop}.
While numerous mechanisms exist for PBH formation (see Refs.~\cite{Sasaki:2018dmp,Carr:2020gox} for a recent review), most assume a spherically symmetric collapse, leading to a non-rotating BH. However, a significant initial spin (Kerr BH) can have profound phenomenological consequences, affecting merger rates, GW emission, superradiance~\cite{Brito:2015oca}, and Hawking evaporation~\cite{Arbey:2019jmj}. Recent applications of TTT to PBH formation in a radiation-dominated universe yield very small spins, $\chi \sim 10^{-3}-10^{-2}$~\cite{DeLuca:2019buf,Mirbabayi:2019uph,DeLuca:2020bjf}, which can be enhanced in a matter-dominated phase if it lasts sufficiently long~\cite{Harada:2017fjm}.

%==========================================================================

In this {\it Letter}, we propose a new, fundamental mechanism for generating cosmological angular momentum. Our approach links the spin of collapsed objects directly to the quantum origin of structure. We consider a spectator complex scalar field, possessing a global U(1) symmetry (e.g., the Peccei-Quinn symmetry), present during inflation. The key insight is that the internal angular momentum of this field—its globally conserved U(1) charge—can be converted into spatial angular momentum.

Our mechanism is rooted in cosmic inflation, during which quantum fluctuations of the field are stretched to superhorizon scales, imprinting spatial perturbations in both its radial and angular components. 
After inflation, if the inflationary dynamics drives the radial mode of the rotation field toward large-field regions, the angular mode can commence rotational motion in field space, storing a large amount of internal angular momentum (charge density)~\cite{Affleck:1984fy,Kuzmin:1985mm,Co:2019wyp}.
This rotation is triggered by higher-dimensional operators that explicitly break the U(1) symmetry, providing an initial ``angular kick'' – a feature generic in UV completions like quantum gravity~\cite{Rai:1992xw,Banks:2010zn,Harlow:2018tng,Witten:2017hdv} where global symmetries are expected to be broken.

When a perturbation mode re-enters the horizon, its inherent spatial gradients couple to the large background charge density. This coupling generates a non-zero momentum density in the energy-momentum tensor, effectively converting the internal field rotation into a bulk flow of energy. If this perturbed region subsequently undergoes gravitational collapse, and if the collapse is non-spherical, this momentum flow manifests as a net spatial angular momentum. Consequently, a PBH formed from such a collapse is born with an intrinsic spin.
If a considerable fraction of DM is composed of the rotation field, the above mechanism may also be applicable to the formation of DM halos.

%=============================================================

\textbf{Field internal angular momentum.}
Consider a theory with a global U(1) symmetry, under which a complex scalar field $P$ transforms as: 
\begin{equation}
    P \to e^{i\alpha} P,\quad P^* \to e^{-i\alpha} P^*,
\end{equation}
where $\alpha$ is an angle parameter.
The Lagrangian is
\begin{equation}
\mathcal{L} = \partial_\mu P \,\partial^\mu P^* - V(|P|^2).
\label{eq:lagrangian_PQ}
\end{equation}
with $V$ depending only on $|P|^2$, ensuring U(1) invariance.
Noether's theorem yields the conserved current associated with this symmetry:
\begin{equation}
j^\mu = i\big( P\,\partial^\mu P^* - P^*\,\partial^\mu P \big).
\label{eq:noether_current_PQ}
\end{equation}
% where we have absorbed the constant $\alpha$ into the definition of the current.
We parameterize the field via a radial mode $S$ and an angular mode $\theta=a/v_s$, where $a$ denotes the axion field and $v_s$ is the vacuum expectation value of $S$:
\begin{equation}\label{eq:field_decomp}
    P = \frac{1}{\sqrt{2}} S\, e^{i\theta},
\end{equation}
The conserved charge density, i.e., the internal angular momentum density, is given by the time component of this current:
\begin{equation}
n_c \equiv j^0 = i\big( P\,\dot{P}^* - P^*\,\dot{P} \big)= S^2 \dot{\theta},
\label{eq:n_PQ_def}
\end{equation}
where the dot denotes the cosmic time derivative.

%===================================================================
% \section{Energy-Momentum Tensor}
\textbf{Energy-momentum tensor.}
For a complex scalar field \(P\), the energy-momentum tensor is
\begin{equation}
T_{\mu\nu} = \partial_\mu P^* \partial_\nu P + \partial_\nu P^* \partial_\mu P
- g_{\mu\nu} \left( g^{\alpha\beta} \partial_\alpha P^* \partial_\beta P - V(P) \right).
\end{equation}
For a perfectly homogeneous rotating field, all spatial derivatives vanish identically. 
In this limit, solely the time component of the energy-momentum tensor is nonvanishing: \(\rho = T_{00} = \frac12 \dot{\theta}^2 S^2 + V(S)\).
The momentum density vanishes trivially, with \(T_{0i} = 0\), as a direct consequence of the absence of spatial gradients; correspondingly, the spatial angular momentum is zero in this configuration.

Scalar perturbations generated during inflation disrupt the spatial homogeneity of the field, giving rise to spatial gradients that source the corresponding spatial momentum.
It is precisely this spatial inhomogeneity that mediates the conversion of intrinsic field rotation into spatial angular momentum.
To analyze a bounded region containing such field perturbations, we decompose
the field into background and perturbed parts:
\begin{equation}
    S = S_0(t) + \delta S(t,\mathbf{x})~~\text{and}~~ \theta = \theta_0(t) + \delta\theta(t,\mathbf{x}),
\end{equation}
where background quantities depend only on cosmic time $t$, while perturbations depend on both cosmic time and physical coordinates $\mathbf{x}$.
Substituting into $T_{0i}$ and expanding to first order in perturbations yields the linearized momentum density:
\begin{equation}\label{eq:T0i_linear}
T_{0i} \approx \dot{S}_0 \partial_i \delta S + S_0^2 \dot{\theta}_0 \partial_i \delta\theta.
\end{equation}
The factor $S_0^2 \dot{\theta}_0$ is precisely the background U(1) charge density. Crucially, the spatial gradients of the perturbations, $\partial_i \delta S$ and $\partial_i \delta\theta$, source a non-zero momentum density. The strength of this source is amplified by the large background charge.

In the scenarios of interest for generating large angular momentum, the radial mode is heavy and sits near the bottom of a steep potential, or its evolution is slow. We can thus neglect the first term in Eq.~\eqref{eq:T0i_linear}. This is quantitatively justified by comparing the two terms for a superhorizon mode with wavenumber $k \sim H$. 
Assuming a potential \(V(S_0)\simeq \frac{1}{2}m_S^2S_0^2\) for \(S_0\gg v_s\) during the matter-like rotation phase~\cite{Co:2021lkc}, circular orbit solutions require vanishing radial acceleration—i.e., the centrifugal force must balance the potential gradient.
This gives \(\dot{\theta}^2S_0\simeq m_S^2S_0\), leading to \(\dot{\theta}_0 \simeq m_S \gg H\).
Since \(n_c^0\) is conserved in the comoving volume, we have \(S_0 \propto a^{-3/2}\) and \(\dot{S}_0/S_0=-3H/2\).
Using the typical amplitude of quantum fluctuations during inflation (\(\delta S \sim H_{\rm inf}/(2\pi)\) and \(\delta\theta \sim H_{\rm inf}/(2\pi S_0)\)), we compute the ratio:
\begin{equation}
\frac{|\dot{S}_0 \partial_i \delta S|}{|S_0^2 \dot{\theta}_0 \partial_i \delta\theta|} \sim \frac{|\dot{S}_0| \cdot \delta S}{S_0^2 \dot{\theta}_0 \cdot \delta\theta} \sim \frac{|\dot{S}_0|}{S_0 \dot{\theta}_0} \sim \frac{3H}{2m_S} \ll 1.
\end{equation}
Hence, the dominant contribution to the spatial momentum arises from the gradient of the phase perturbation, weighted by the large internal charge density \(n_c^0 \,\partial_i \delta\theta\).
This constitutes the central dynamical equation underlying our proposed mechanism: the intrinsic rotational energy of the field is converted into a bulk spatial momentum flow via spatial variations in the field phase seeded during inflation.

Notably, upon entering the nonlinear regime of gravitational collapse, where the amplitude of perturbations grows sufficiently large, second-order contributions can no longer be neglected or truncated.
In this regime, second-order terms within the energy-momentum tensor component \(T_{0i}\) — such as the term \(2 S_0 \delta S \dot{\theta}_0 \partial_i (\delta \theta)\) — begin to generate additional momentum density.

%================================================================

% \section{Angular Momentum}
\textbf{Spatial angular momentum.}
The total spatial angular momentum of a region of volume \(V\) is given by
\begin{equation}
J_{ij} = \int_V d^3x \left( x_i T_{0j} - x_j T_{0i} \right).
\end{equation}
Substituting the expression~\eqref{eq:T0i_linear} for \(T_{0i}\), using integration by parts and assuming vanishing boundary contributions (valid for a collapsing isolated region), we obtain:
\begin{equation}\label{eq:Jij}
J_{ij} \approx -S_0^2 \dot{\theta}_0
\int_V d^3x \delta\theta \cdot (x_i \partial_j - x_j \partial_i).
% + \text{boundary terms}.
\end{equation}

%==========================================================================
% \subsection{Spherical collapsing}
{\it Spherical collapse.}
% Let us first consider the case of spherical collapse. 
Working in spherical coordinates, we take a spherical collapsing region of radius \(R\) bounded by the surface \(\partial V\). The total angular momentum vector is given by the following surface integral:
\begin{equation} 
J_i = S_0^2 \dot{\theta}_0 \, \epsilon_{ijk} \oint_{\partial V} dS \, x_j n_k \, \delta\theta(\Omega), 
\end{equation}
where \(\mathbf{n}\) denotes the outward unit normal vector of the spherical boundary, \(\mathbf{x} = R \mathbf{n}\), and the surface element is defined as \(dS = R^2 d\Omega\). Substitution yields the simplified expression:
\begin{equation}\label{eq:Ji1} 
J_i = S_0^2 \dot{\theta}_0 R^3 \, \epsilon_{ijk} \int d\Omega \, n_j n_k \, \delta\theta(\Omega),
\end{equation}
where \(\delta\theta(\Omega)\) represents the angular distribution of the phase perturbation across the spherical surface.

The three components of the unit direction vector \(\mathbf{n}\) can be written as linear combinations of the \(Y_{1m}\) spherical harmonics. Correspondingly, the product \(n_j n_k\) forms a linear combination of \(Y_{1m_1}Y_{1m_2}\), which further decomposes into spherical harmonics of orbital angular momentum \(l=0\) and \(l=2\). Explicitly, we have the identity:
\begin{equation}\label{eq:nnY} 
n_j n_k - \frac{1}{3}\delta_{jk} = \sum_{m=-2}^{2} c_{jk}^{(2m)} Y_{2m}(\Omega), 
\end{equation}
where \(\delta_{jk}\) is the Kronecker delta function and \(c_{ij}^{(2m)}\) are angle-independent constant coefficients; for a fixed \(m\), \(c_{ij}^{(2m)}\) corresponds to a symmetric traceless tensor, whose values can be derived from Clebsch–Gordan coefficients. We next expand the phase perturbation in terms of spherical harmonics:
\begin{equation}\label{eq:dtheta_expansion} 
\delta\theta(\Omega) = \sum_{l=0}^{\infty}\sum_{m=-l}^{l} a_{lm} Y_{lm}(\Omega). 
\end{equation}
Substituting Eqs.~\eqref{eq:nnY} and~\eqref{eq:dtheta_expansion} into Eq.~\eqref{eq:Ji1}, and noting that \(\epsilon_{kij}\delta_{ij} =  0 = \epsilon_{kij} c_{ij}^{(2m)}\) — a direct consequence of the symmetry of \(\delta_{ij}\) and \(c_{ij}^{(2m)}\) versus the antisymmetry of the Levi-Civita tensor \(\epsilon_{kij}\) with respect to indices \(i,j\) — we find that \(J_i = 0\). This implies that, for spherically symmetric collapse, the total angular momentum vanishes identically for arbitrary angular distributions of the field phase perturbation.
We thus conclude that PBHs formed via purely spherical collapse acquire no net angular momentum. 

{\it Ellipsoidal collapse.}
We now extend our analysis to non-spherically symmetric collapsing configurations, focusing on the ellipsoidal collapse as a representative case.
For collapse within an ellipsoidal region, the geometric anisotropy implies that the normal vector $n_j$ is no longer parallel to the position vector $x_j$.
Consequently, the geometric factor $\epsilon_{kij} x_i n_j$ is dominated by $l=2$ (quadrupolar) contributions in its spherical harmonic expansion.
Only angular perturbation modes with $l\geq 2$ can therefore couple to this geometric factor and yield a non-vanishing angular momentum.

We parameterize the angular perturbation $\delta\theta$ using a symmetric, traceless quadrupole tensor $Q_{ab}$ in Cartesian coordinates:
\begin{equation}\label{eq:delta_theta}
\delta\theta(\mathbf{x}) = Q_{ab} x^a x^b,
\end{equation}
with summation over repeated indices.
The corresponding gradient reads
\begin{equation}
\partial_j \delta\theta = 2 Q_{j a} x^a.
\end{equation}
Substituting into Eq.~\eqref{eq:Jij}, we obtain
\begin{equation}
J_k = 2 S_0^2 \dot{\theta}_0 \, \epsilon_{kij} Q_{j a} \int_V x_i x^a \, d^3x
= 2 S_0^2 \dot{\theta}_0 \, \epsilon_{kij} Q_{j a} I_{ia},
\end{equation}
where we define the second-moment tensor of the volume
\begin{equation}\label{eq:Iia}
I_{ia} = \int_V x_i x_a \, d^3x.
\end{equation}

Consider an ellipsoid aligned with the coordinate axes, with semi-axes $a,b,c$:
\begin{equation}
V: \frac{x^2}{a^2} + \frac{y^2}{b^2} + \frac{z^2}{c^2} \le 1.
\end{equation}
Its volume is $V_0 = \frac{4\pi}{3}abc$.
From Eq.~\eqref{eq:Iia}, the second-moment tensor is diagonal,
\begin{equation}
I_{ia} = \tilde{I}_{ia} \delta_{ia},
\end{equation}
with the diagonal factors \(\tilde{I}_{xx} = \frac{4\pi}{15}a^3 b c\), \(\tilde{I}_{yy} = \frac{4\pi}{15}a b^3 c\), and \(\tilde{I}_{zz} = \frac{4\pi}{15}a b c^3\).
While all off-diagonal components vanish, $I_{xy}=I_{xz}=I_{yz}=0$.
The angular momentum then simplifies to
\begin{equation}\label{eq:angular_momentum}
J_i = 2 S_0^2 \dot{\theta}_0 \, Q_{jk} (\tilde{I}_{jj} - \tilde{I}_{kk}),
\end{equation}
where $i,j,k$ label the Cartesian coordinates.
We observe that only off-diagonal quadrupole components (corresponding to $l=2$, $m=\pm2$ modes) contribute to $J$.
Furthermore, the angular momentum vanishes identically in the spherical limit $a=b=c$.

%=============================================================

% \section{Perturbations from inflation}

\textbf{Primordial perturbations.}
We assume the scalar fields are massless (or $m_S\ll H_{\rm inf}$) during inflation. 
For a massless field \(\phi\), quantum fluctuations amplified by inflation are nearly scale-invariant.
The power spectrum is defined as:
\begin{equation}\label{eq:PP1}
\langle \delta\phi_{\mathbf{k}} \delta\phi_{\mathbf{k}'}^* \rangle = (2\pi)^3 \delta^3(\mathbf{k} - \mathbf{k}') P_{\delta\phi}(k),
\end{equation}
where the Fourier component is:
\begin{equation}
\delta\phi(\mathbf{x}) = \int \frac{d^3k}{(2\pi)^3} e^{i\mathbf{k}\cdot\mathbf{x}} \delta\phi_{\mathbf{k}}.
\end{equation}
For a massless field, standard inflationary models give:
\begin{equation}\label{eq:power}
P_{\delta\phi}(k) = \frac{2\pi^2}{k^3} \Delta_{\delta\phi}^2(k),\quad \Delta_{\delta\phi}^2(k) = \left( \frac{H_{\text{inf}}}{2\pi} \right)^2 \left( \frac{k}{k_*} \right)^{n_{s}-1},
\end{equation}
where $H_{\text{inf}}$ is the Hubble parameter during inflation, $k_*$ is a reference scale (typically the CMB scale), and $n_{s} \approx 1$ is the spectral index (nearly scale-invariant).

Decompose $P$ into two real scalar fields, \(P=(\phi_1 + i\phi_2)/\sqrt{2}\),
where $\phi_1$ and $\phi_2$ are independent free real scalar fields with mass $m_S\ll H_{\text{inf}}$.
Relating $\phi_1, \phi_2$ to $S$ and $\theta$ for a background field that can be aligned along the real axis ($\theta_0=0$), we have $\delta S = \delta\phi_1$ and $\delta\theta = \delta\phi_2 / S_0$.
Consequently, the power spectra for the perturbations at the end of inflation are:
\begin{equation}
    \Delta_{\delta S}^2(k) = \frac{k^3}{2\pi^2} P_{\delta S}(k) = \left( \frac{H_{\text{inf}}}{2\pi} \right)^2\left(\frac{k}{k_*} \right)^{n_{s}-1},
\end{equation}
\begin{equation}\label{eq:dtheta_power}
    \Delta_{\delta\theta}^2(k) = \frac{k^3}{2\pi^2} P_{\delta\theta}(k) = \frac{1}{S_0^2} \left( \frac{H_{\text{inf}}}{2\pi} \right)^2\left(\frac{k}{k_*} \right)^{n_{s}-1}.
\end{equation}
The power spectra are initialized at the end of inflation when all modes of interest have exited the horizon.
These perturbations remain frozen on superhorizon scales until a mode with wavenumber $k$ re-enters the horizon at a later epoch.

%==========================================================================

% \section{Estimation of the spin}
\textbf{Estimation of PBH spin.}
We now estimate the angular momentum of a PBH formed from the collapse of a field overdensity during the matter-like rotation phase~\cite{Co:2021lkc,Gouttenoire:2021wzu}.
From Eq.~\eqref{eq:delta_theta}, the quadrupole tensor is \(Q_{jk} = \frac{1}{2} \partial_j\partial_k \delta\theta\).
The typical amplitude of the field perturbation during inflation is \(\delta\theta \sim \Delta_{\delta\theta}\).
Since \(Q_{jk}\) corresponds to a second spatial derivative, we estimate it as:
\begin{equation}
Q_{jk} \sim k^2 \delta\theta \sim H^2\Delta_{\delta\theta},
\label{eq:Q_jk_H}
\end{equation}
where we use \(k \sim H\) (with \(H\) denoting the Hubble parameter at PBH formation) for horizon-crossing perturbations.
The mass of a horizon-sized BH is estimated as:
\begin{equation}
M \sim \rho_{\theta}R^3\sim \left(\frac{H^2}{G} \right)H^{-3} \sim \frac{M_{\text{Pl}}^2}{H},
\label{eq:M_H}
\end{equation}
where \(M_{\text{Pl}}=2.43\times 10^{18}\ \text{GeV}\) is the reduced Planck mass.
The energy density of the field rotation is \(\rho_\theta \sim \dot{\theta}_0^2 S_0^2\).
When the field rotation energy dominates the Universe, the Friedmann equation gives \(\rho_\theta \sim H^2 M_{\text{Pl}}^2\), leading to:
\begin{equation}
\dot{\theta}_0 S_0 \sim H M_{\text{Pl}}.
\label{eq:thetaS_H_MPl}
\end{equation}

For a uniform ellipsoidal region, the semi-axes satisfy \(a, b, c \sim R\), with \(a \approx R(1+\varepsilon_a)\) (and analogous expressions for \(b,c\)).
We set \(b = R(1+\varepsilon)\) and \(c = R(1-\varepsilon)\) to preserve the volume (\(\sim abc \sim R^3\)), yielding the moment of inertia difference:
\begin{equation}\label{eq:I_appro}
\tilde{I}_{yy} - \tilde{I}_{zz} = \frac{4\pi}{15}abc(b^2 - c^2) \sim \frac{4\pi}{15}R^5 \cdot 4\varepsilon \sim \epsilon H^{-5},
\end{equation}
where \(\varepsilon\) is the ellipticity parameter and \(\epsilon=16\pi \varepsilon/15\sim 0.1\) denotes the ellipsoidal form factor.
Substituting Eqs.~\eqref{eq:Q_jk_H}, \eqref{eq:I_appro}, and~\eqref{eq:thetaS_H_MPl} into Eq.~\eqref{eq:angular_momentum}, we obtain:
\begin{equation}
J \sim S_0^2 \dot{\theta}_0 \cdot H^2\Delta_{\delta\theta} \cdot \epsilon H^{-5}\sim \epsilon M_{\text{Pl}}\Delta_{\delta S}H^{-2}.
\label{eq:J_final}
\end{equation}

We introduce the dimensionless spin parameter \(\chi = \frac{J}{G M^2} = \frac{J M_{\text{Pl}}^2}{M^2}\).
Using Eqs.~\eqref{eq:J_final} and~\eqref{eq:M_H}, we find:
\begin{equation}
\chi \sim \epsilon \frac{\Delta_{\delta S}}{M_{\text{Pl}}}\sim \frac{\epsilon}{2\pi}\frac{H_{\text{inf}}}{M_{\text{Pl}}},
\label{eq:chi_final}
\end{equation}
where the second approximation invokes the scale-invariant power spectrum.
This shows that the perturbation power spectrum determines the initial PBH spin during inflation.
For the scale-invariant spectrum and taking the CMB-constrained upper bound \(H_{\mathrm{inf}} \sim 10^{14}\ \text{GeV}\) for the inflationary Hubble rate, the typical PBH spin is estimated to be \(\sim 10^{-5}\), which is negligibly small.

Both the PBH spin and abundance can be significantly enhanced by considering a scale-dependent power spectrum—e.g., in multi-scalar inflation scenarios~\cite{Silk:1986vc,Randall:1995dj,GarciaBellido:1996mdl}.
Current CMB and large-scale structure observations constrain the primordial curvature power spectrum to \(\sim 10^{-9}\) on large scales ($\lesssim \mathrm{Mpc}^{-1}$), implying negligible PBH formation from inflationary density perturbations.
To produce a detectable PBH abundance (consistent with LVK observations), the primordial curvature power spectrum must be enhanced to \(\sim 10^{-2}-10^{-1}\) on small scales~\cite{Sasaki:2018dmp,Byrnes:2018clq}.
In this case, the PBH spin can be boosted to \(\chi\sim 0.1-1\)—far larger than the spin generated via the TTT mechanism.

% \section{Conclusion}
\textbf{Discussion.}
We have proposed a novel mechanism that traces the fundamental origin of cosmic spatial angular momentum to the intrinsic phase-space rotation of the scalar field.
This mechanism necessitates two critical physical prerequisites: sufficiently large radial field amplitudes to sustain substantial field rotation, and finite field perturbations to explicitly break the spatial homogeneity of the background field configuration. Both of these requisite conditions are naturally fulfilled within the inflationary cosmological framework.
The angular field component attains an initial rotational speed through torque sourced by a higher-dimensional symmetry-breaking operator.
As a perturbation mode re-enters the cosmological horizon, the spatial gradients of such perturbations couple to the background charge density, generating a non-vanishing momentum density that manifests as a coherent bulk flow. During the subsequent gravitational collapse, this bulk flow is subsequently converted into a net macroscopic spatial angular momentum.
We demonstrated that only non-spherical (e.g., ellipsoidal) collapse configurations enable the quadrupolar modes of the phase perturbation to couple to the spacetime geometry and yield a non-vanishing spin.

We find that the initial PBH spin generated by this mechanism is determined by the primordial power spectrum. If the PBH abundance is sufficient for detection by LVK, the PBH spin can be boosted to $\chi\sim 0.1-1$, far exceeding the spin values predicted by TTT.
Notably, this mechanism predicts a spin distribution intimately linked to the small-scale power spectrum, which may exhibit a bimodal or broad shape with high-spin tails.
Studies on gas accretion onto BHs have established an upper limit on the dimensionless spin $\chi_{\rm lim}\simeq 0.998$ for astrophysical BHs~\cite{Thorne:1974ve,Kesden:2009ds,Sadowski:2011ka,Arbey:2019jmj}.
A high-spin PBH violating the Thorne limit would thus provide distinctive evidence for the quantum origin of the spatial angular momentum.

Most of the existing literature focuses on spinless PBHs. High-spin PBHs could give rise to a wealth of additional intriguing physical phenomena. Firstly, high spin magnitudes can enhance Hawking radiation~\cite{Murata:2006pt,Dai:2010xp}, thereby amplifying the corresponding PBH observational signatures imprinted on big bang nucleosynthesis and the CMB~\cite{Acharya:2020jbv}. Furthermore, PBH spin provides a reservoir of rotational energy that can drive BH superradiance; the induced signals from this process, including enhanced DM signatures and GWs~\cite{Baryakhtar:2017ngi,Bernal:2022oha}, may be detectable with state-of-the-art terrestrial experiments. Additionally, PBH spin can modify the PBH merger rate and the resulting GW waveforms, predictions that are directly testable via LVK observations~\cite{Gow:2019pok}.
In summary, the mechanism proposed in this {\it Letter} establishes a testable physical connection between quantum theory, inflationary cosmology, and the observable properties of PBHs. The LVK experiment~\cite{LIGOScientific:2025slb}, together with upcoming GW observatories-such as LISA~\cite{Berti:2005ys}, Taiji~\cite{Hu:2017mde}, TianQin~\cite{TianQin:2015yph}, and the Einstein Telescope~\cite{Branchesi:2023mws}-paired with advanced axion detection experiments~\cite{ADMX:2018gho}, opens up highly promising avenues to experimentally probe these distinctive phenomena.

%==========================================================================

% \section*{Acknowledgements}
\textbf{Acknowledgements.}
BQL is supported in part by the National Natural Science Foundation of China under Grant No.~12405058 
and by the Zhejiang Provincial Natural Science Foundation of China under Grant No.~LQ23A050002.

\bibliography{reference} 
%==========================================================================
 
% \clearpage
% \newpage
% \maketitle
% \onecolumngrid
% \begin{center}
% \textbf{\large Clockwork axion footprint on nano-hertz stochastic gravitational wave background} \\ 
% \vspace{0.05in}
% { \it \large Supplemental Material}\\ 
% {By Bo-Qiang Lu, Cheng-Wei Chiang, and Tianjun Li}
% \vspace{0.05in}
% \end{center}
% \onecolumngrid
% %%%%%%%%%% Merge with Supplemental material %%%%%%%%%%
% \setcounter{equation}{0}
% \setcounter{figure}{0}
% \setcounter{table}{0}
% \setcounter{section}{0}
% \setcounter{page}{1}
% \makeatletter
% \renewcommand{\theequation}{S\arabic{equation}}
% \renewcommand{\thefigure}{S\arabic{figure}}
% \renewcommand{\thetable}{S\arabic{table}}

%==========================================================================
% \clearpage
% \newpage
% \maketitle
% \onecolumngrid
% \appendix
%==========================================================================

\end{document}